\documentclass[nofootinbib]{revtex4}
\usepackage{graphicx}
\usepackage{latexsym}
\def\be{\begin{equation}}
\def\ee{\end{equation}}
\def\bea{\begin{eqnarray}}
\def\eea{\end{eqnarray}}

\begin{document}

\hfill USTC-ICTS-15-04

\title{Conformal invariant cosmological perturbations via the covariant approach}

\author{Mingzhe Li$^{1,2}$}
\email{limz@ustc.edu.cn}
\author{Yicen Mou$^{1}$}
\affiliation{$^{1}$Interdisciplinary Center for Theoretical Study, University of Science and Technology of China, Hefei, Anhui 230026, China}
\affiliation{$^{2}$State Key Laboratory of Theoretical Physics, Institute of Theoretical Physics, Chinese Academy of Sciences, Beijing 100190, China}

\begin{abstract}

It is known that some cosmological perturbations are conformal invariant. This facilitates the studies of perturbations within some gravitational theories alternative to general relativity, for example the scalar-tensor theory, because it is possible to do equivalent analysis in a certain frame in which the perturbation equations are simpler. In this paper we revisit the problem of conformal invariances of cosmological perturbations in terms of the covariant approach in which the perturbation variables have clear geometric and physical meanings. We show that with this approach the conformal invariant perturbations are easily identified. 

\end{abstract}

\maketitle

\hskip 1.6cm PACS number(s): 98.80.-k \vskip 0.4cm

\section{Introduction}

Cosmological perturbation theory constitutes the cornerstone of our current understanding of the origin, evolution and formation of the cosmic large-scale structures.
This theory starts from the splitting of the spacetime of the physical universe into a fictitious Friedmann-Lema\^{i}tre-Robertson-Walker (FLRW) background and small perturbations around it.
An important problem one should carefully deal with is the gauge issue which reflects the arbitrariness of the correspondence between the spacetime of the background and that of the real universe.  
A convenient way  to circumvent this problem is by focusing on the gauge invariant variables.  In the standard approach (or coordinate approach) to the cosmological perturbation theory,  gauge invariant perturbations can be constructed 
(usually in a non-geometric way) as the combinations of gauge dependent metric and matter perturbations in a given coordinate system,
as first done by Bardeen \cite{Bardeen:1980kt} and reviewed in Refs. \cite{Kodama:1985bj,Mukhanov:1990me} for the linear perturbation theory.   
In the covariant approach (or geometric approach)  developed in \cite{Ellis:1989jt,Ellis:1989ju,Bruni:1992dg,Dunsby:1991xk}
and based on earlier works by Hawking \cite{Hawking:1966qi} and Ellis \cite{Ellis:1971pg}, all the variables are covariantly defined. Crucial variables are those having vanishing values in the background FLRW universe. They themselves are perturbations in the real universe, and according to the Stewart-Walker Lemma \cite{Stewart:1974uz} they are automatically gauge invariant\footnote{There are other two possibilities to obtain covariant and gauge invariant quantities according to the Stewart-Walker Lemma: if the background part of a covariantly defined variable is either a constant scalar or a linear combination of products of Kronecker deltas, its perturbation is gauge invariant. To our knowledges, no important quantity in the cosmological perturbation theory is constructed from these two possibilities.}. One of examples is the Weyl tensor which vanishes in the FLRW universe because the FLRW spacetime is conformally flat. Hence all of the non-zero components of the Weyl tensor in the inhomogeneous universe must be gauge invariant. 
The advantage of this approach is that the gauge invariant variables have clear geometric and physical meanings through the covariant definitions. Some futher discussions about this approach can be found, e.g., in \cite{Vitenti:2013hda,Osano:2006ew}.

Besides the gauge issue, in recent years there are lots of interests in investigating the changes of cosmological perturbations under the conformal transformation or Weyl rescaling
\be\label{conformal}
\tilde{g}_{ab}=\Omega^2(x) g_{ab}~,
\ee
where $a,b=0,1,2,3$ and the conformal factor $\Omega(x)$ is arbitrary real function of the coordinates and will be assumed to be positive in this paper. 
The conformal transformation, keeping the coordinate system fixed, will change from one frame to another. It is frequently used in the scalar-tensor theory of gravity, where in the Jordan frame the action of gravity contains a non-minimal coupling of a scalar field to the curvature scalar but the matter is minimally coupled to the metric. After a conformal transformation to the Einstein frame, the action of gravity becomes the Einstein-Hilbert one but the matter has a non-minimal coupling to the scalar field. These theories have no conformal symmetry and equations of motions have quite different forms in different frames, so sometimes the physical equivalence of different frames were questioned. However, the conformal transformations, considered as passive transformations, are essentially local field redefinitions of the same physical system and should not change the physics. So different frames should be physically equivalent and the observables are expected to be frame independent. To demonstrate this, it is crucial to find out the variables which are frame independent and correspond to what we measure in the experiments and observations. 
It was found \cite{Catena:2006bd,Deruelle:2010ht,Chiba:2013mha} that the observables in the FLRW background, such as the redshift and the redshift-magnitude relation, are conformal invariant though the background evolution itseflt depends on the frame. Some perturbations were shown \cite{Catena:2006bd,Chiba:2013mha,Gong,Chiba,Prokopec,Kubota:2011re} to be conformal invariant at the linear and non-linear order in terms of the coordinate approach. Especially the famous comoving curvature perturbation $\zeta$ \cite{Bardeen:1983qw} in single scalar field inflation models was found to be frame independent up to the full non-linear order \cite{Gong}. This is important for some early universe models which account for the primordial perturbations. In these models, such as the Higgs inflation \cite{Bezrukov:2007ep}, non-minimal couplings between gravity and a scalar field are involved and their predictions on the primordial perturbations can be calculated in the more familiar Einstein frame.  Some implications of the equivalence between different frames on the early universe were discussed in Refs. \cite{Li,Piao,Qiu:2012ia,Domenech:2015qoa}, other discussions can be found in Refs. \cite{other1,other2}.

Up to now all the studies on the conformal invariances or frame independences of cosmological perturbations were based on the coordinate approach. As mentioned above the covariant approach has some advantages to obtain the gauge invariant perturbations. With clear geometric meanings this approach is also expected to have advantages to show up the properties of perturbations under the conformal transformation. A simple example is the aforementioned Weyl tensor. It is gauge invariant according to the Stewart-Walker Lemma, and it is well known that it is also conformal invariant. The purpose of this paper is to investigate how the gauge invariant perturbations obtained via the covariant approach change under the conformal transformation, and we will show that the covariant approach provides a simple way to identify the conformal invariant cosmological perturbations. 
This paper is organized as follows: in Section II, we will briefly review the covariant approach; Some conformal invariant perturbations will be identified in Section III; The links to the coordinate approach will be presented in Section IV; In Section V we will discuss the gauge issue when expanding the covariant variables to higher order perturbations; Section VI is our conclusion.

\section{Brief review on the covariant approach}

In the covariant approach \cite{Ellis:1989jt,Ellis:1989ju,Bruni:1992dg,Dunsby:1991xk} we first choose a preferred family of world lines which represent the motion of typical observers (fundamental observers) in the universe. The tangent vector of the flow lines $u^a=dx^a/d\lambda$ is the four-velocity of the fundamental observers, where $\lambda$ is the proper time along the flow lines. The four-velocity is timelike, future-directed and unit, $u_a u^a=-1$. Then we introduce the projection tensor into the tangent three-space orthogonal to $u^a$
\be
h_{ab}=g_{ab}+u_au_b~,~{\rm with}~h^a_{~b}h^b_{~c}=h^a_{~c}~,~h_a^{~b}u_b=0~. 
\ee
It is useful to decompose the first covariant derivative of the four-velocity as
\be
\nabla_b u_a=\omega_{ab}+\sigma_{ab}+\frac{1}{3}\Theta h_{ab}-a_au_b~,
\ee
where $\omega_{ab}$ is the antisymmetric vorticity tensor with $\omega_{ab}u^b=0$, $\sigma_{ab}$ is the symmetric shear tensor with $\sigma_{ab}u^b=0$ and $\sigma^a_{~a}=0$, $\Theta\equiv \nabla_a u^a$ is the local expansion rate, and $a_a= \dot u_a\equiv u^b\nabla_b u_a$ is the acceleration vector with $a_a u^a=0$. The vorticity and shear magnitudes are defined by $\omega^2=(1/2)\omega_{ab}\omega^{ab}$, $\sigma^2=(1/2)\sigma_{ab}\sigma^{ab}$. 
If the vorticity $\omega_{ab}$ vanishes, there exists a family of hypersurfaces $\Sigma={\rm constant}$ to which $u^a$ is orthogonal everywhere, and $\sigma_{ab}+(1/3)\Theta h_{ab}$ is the second fundamental form of the hypersurface, i.e., its extrinsic curvature. In general case, $\omega_{ab}\neq 0$, these hypersurfaces cannot be defined.
It is also useful to introduce the local scale factor $S=e^{\alpha}$, here $\alpha$ is the integration of $\Theta$ along the flow lines with respect to the proper time
\be\label{alpha}
\alpha\equiv \frac{1}{3}\int d\lambda \Theta~,
\ee
which is defined up to an integration constant.
The matter sector is described by its energy momentum tensor. For single perfect fluid, the fluid velocity is identified with the four-velocity of the fundamental observers $u^a$ and its energy momentum tensor is decomposed as
\be
T_{ab}=\rho u_au_b+ph_{ab}~,
\ee
where $\rho=T_{ab}u^au^b$ is proper density, $p=(1/3)h^{ab}T_{ab}$ is the pressure.

In the FLRW universe, the shear $\sigma_{ab}$, vorticity $\omega_{ab}$ and acceleration $a_a$ vanish, so themselves are gauge invariant perturbations. Other gauge invariant perturbations can be obtained from the spatial gradients 
of various scalar quantities, such as \cite{Ellis:1989jt}
\be
X_a=D_a \rho~,~Y_a=D_a p~,~Z_a=D_a\Theta~,
\ee
and the one introduced in Refs. \cite{Langlois:2005ii,Langlois:2005qp}
\be
W_a= D_a \alpha~,
\ee
 where the derivative $D_a\equiv h_a^{~b}\nabla_b$ is the projection of the covariant derivative into the tangent three-space.

In addition, we have some curvature variables which vanish in the background FLRW spacetime. One of which is the Weyl tensor, but it is more convenient to use its electric and magnetic parts
\be
E_{ab}=C_{acbd}u^c u^d~,~H_{ab}={1\over 2} C_{aecd}u^e \eta^{cd}_{~~~bf}u^f~,
\ee
where $\eta^{cd}_{~~~bf}$ is the totally anti-symmetric tensor, with $\eta_{0123}=\sqrt{-g}\equiv \sqrt{-|{\rm det}g_{ab}|}$. 
Moreover, one has the three-curvature at each point
\bea\label{codazzi}
\hat R=R+2 R_{ab}u^au^b-2(\frac{\Theta^2}{3}+\omega^2-\sigma^2)~.
\eea
In the case of hypersurface-orthogonal, the vorticity vanishes, $\hat R$ is the curvature scalar of the hypersurface $\Sigma={\rm constant}$ and Eq. (\ref{codazzi}) is the contracted Gauss-Codazzi equation. The three-curvature scalar only vanishes in the spatial-flat FLRW universe, in general its perturbation $\delta \hat R$ is not gauge invariant. In the literature its associated gauge invariant perturbation is constructed by its spatial gradient in most cases \cite{Bruni:1992dg}
\be
C_a=S^3D_a \hat R~.
\ee
In this paper, however we will instead consider the following quantity
\be\label{gaugecurvature}
\mathcal{C} =S^2 \hat R-6K~,
\ee
which vanishes in FLRW spacetime with any constant spatial curvature $K$, hence is automatically a gauge invariant perturbation.

\section{Conformal invariant perturbations}

The conformal transformation (\ref{conformal}) preserves the normalization of four-velocity
\be
\tilde{g}_{ab}\tilde{u}^a\tilde{u}^b=g_{ab}u^au^b=-1~,
\ee
so one has the relation $\tilde{u}^a=\Omega^{-1} u^a$. This means the proper distance changes by a dilation, $d\tilde{\lambda}=\Omega d\lambda$. 
Keeping this in mind, it is immediately to find the change rules of the kinematical variables
\bea\label{con1}
& &\tilde{\omega}_{ab}=\Omega\omega_{ab}~,~\tilde{\sigma}_{ab}=\Omega\sigma_{ab}~,~{\rm or}~\tilde{\omega}=\Omega^{-1}\omega~,~\tilde{\sigma}=\Omega^{-1}\sigma~,\nonumber\\
& &\tilde{\Theta}=\Omega^{-1}\Theta+3\Omega^{-2}\dot\Omega~,~\tilde{a}_a=a_a+D_a\ln\Omega~,\nonumber\\
& &\tilde{\alpha}=\alpha+\ln \Omega~,~\tilde{S}=\Omega S~,~~
\tilde{W}_a=W_a+ D_a\ln \Omega~.
\eea

Furthermore, for any scalar with conformal weight $w$, $\tilde{\phi}=\Omega^{-w}\phi$, its spatial gradient is transformed as
\be
\tilde{D}_a\tilde{\phi}=D_a\tilde{\phi}=\Omega^{-w}(D_a\phi-w\phi D_a\ln\Omega)~,
\ee
so $\phi^{-1}D_a\phi$ is both gauge and conformal invariant for the scalar field with conformal weight zero. 
The energy-momentum tensor for matter changes as 
\be
\tilde{T}_{ab}=\Omega^{-2}T_{ab}~,
\ee 
this can be seen from the definition of the energy-momentum tensor through the variation, $T^{ab}= (2/ \sqrt{-g})(\delta S_m/\delta g_{ab})$, where $S_m$ is the matter action. From this it is straightforwardly to know that the energy density and pressure of matter have the conformal weight $4$, i.e.,
$\tilde{\rho}=\Omega^{-4}\rho~,~\tilde{p}=\Omega^{-4} p$,
and the perturbations change as
\be\label{con2}
\tilde{X}_a=\Omega^{-4}(X_a-4\rho D_a\ln \Omega)~,~\tilde{Y}_a=\Omega^{-4}(Y_a-4p D_a\ln \Omega)~.
\ee
Hence the following ratios get a shift by conformal transformation, 
\be\label{concurvature}
\frac{\tilde{X}_a}{\tilde{\rho}}=\frac{X_a}{\rho}-4D_a\ln\Omega~,~\frac{\tilde{Y}_a}{\tilde{p}}=\frac{Y_a}{p}-4D_a\ln\Omega~.
\ee

As far as the curvature variables are concerned, we first have the electric and magnetic parts of the Weyl tensor which are conformal invariant,
\be\label{con3}
\tilde{E}_{ab}=E_{ab}~,~\tilde{H}_{ab}=H_{ab}~.
\ee
The change of the three-curvature can be calculated through Eq. (\ref{con1}) and the conformal transformations of the curvature scalar $R$ and Ricci scalar $R_{ab}$, see for example \cite{Carroll:2004st},
\bea
\tilde{R}&=& \Omega^{-2}R-6g^{ab}\Omega^{-3}\nabla_a\nabla_b\Omega~,\nonumber\\
\tilde{R}_{ab}&=& R_{ab}-[2\delta^c_a\delta^d_b+g_{ab}g^{cd}]\Omega^{-1}\nabla_c\nabla_d\Omega+[4\delta^c_a\delta^d_b-g_{ab}g^{cd}]\Omega^{-2}\nabla_c\Omega \nabla_d\Omega~,
\eea
thus Eq. (\ref{codazzi}) gives
\bea
\tilde{\hat R}&=&\tilde{R}+2 \tilde{R}_{ab}\tilde{u}^a\tilde{u}^b-2(\frac{\tilde{\Theta}^2}{3}+\tilde{\omega}^2-\tilde{\sigma}^2)\nonumber\\
&=& \Omega^{-2}  [ \hat R-4\Omega^{-1} (h^{ab}\nabla_a\nabla_b \Omega+ \Theta \dot\Omega)+2 \Omega^{-2} h^{ab}\nabla_a\Omega\nabla_b\Omega]~.
\eea
This implies the gauge invariant variable $\mathcal{C}$ defined in Eq. (\ref{gaugecurvature}) changes as
\be\label{con4}
\tilde{\mathcal{C}}=\mathcal{C}
-S^2[4\Omega^{-1} (h^{ab}\nabla_a\nabla_b \Omega+ \Theta \dot\Omega)-2 \Omega^{-2} h^{ab}\nabla_a\Omega\nabla_b\Omega]~.
\ee

With these formulae (\ref{con1}), (\ref{con2}), (\ref{con3}) and (\ref{con4}), one can find out the gauge invariant perturbations which are also conformal invariant, such as 
\be\label{conformal1}
E_{ab},~H_{ab},~\frac{\omega_{ab}}{S},~\frac{\sigma_{ab}}{S},~W_a-a_a~,~\frac{X_a}{\rho}+4W_a~,~\frac{Y_a}{p}+4W_a~,~\frac{X_a}{\rho}-\frac{Y_a}{p}~.
\ee
The quantities listed above are conformal invariant with any conformal factor $\Omega(x)$.

Special interests arise in the case of hypersurface-orthogonal. These spacelike hypersurfaces are labeled by $\Sigma={\rm constant}$ with $\Sigma$ a scalar function. 
As known from above discussions, quantities like $W_a$, $a_a$, $X_a/\rho$, $Y_a/p$ are not conformal invariant in general, but they are indeed invariant under the subset of conformal transformations in which the conformal factor only depends on $\Sigma$, i.e., 
$\Omega=\Omega(\Sigma)$. Because $u^a$ is orthogonal to the hypersurface $\Sigma={\rm constant}$, it is easy to show that $D_a\ln \Omega(\Sigma)=0$ and $\tilde{W}_a=W_a,~
\tilde{a}_a=a_a,~\tilde{X}_a/\tilde{\rho}=X_a/\rho,~\tilde{Y}_a/\tilde{p}=Y_a/p$. More importantly one can show that in this case
\bea
h^{ab}\nabla_a\nabla_b \Omega+ \Theta \dot\Omega=0~,~h^{ab}\nabla_a\Omega\nabla_b\Omega=0~,
\eea
so from Eq. (\ref{con4}) the variable $\mathcal{C}$ associated with the three-curvature is also invariant under the transformations with $\Omega(\Sigma)$, i.e., $\tilde{\mathcal{C}}=\mathcal{C}$.  
In all, we picked out another kind of conformal invariant perturbations, $W_a$, $a_a$, $X_a/\rho$, $Y_a/p$ and $\mathcal{C}$ which are invariant under a subset of conformal transformations in the special case of hypersurface-orthogonal, 
supplementary to the quantities listed in Eq. (\ref{conformal1}). This is crucial for studies of cosmological perturbations in scalar tensor theories. In these theories the transforms between different frames are realized by the conformal transformations with the factors only depending on the scalar field, $\Omega=\Omega(\phi)$. The four-velocity $u^a$ is conveniently chosen to be normal to the hypersurface $\phi={\rm constant}$. We know from above discussions that besides the perturbations listed in Eq. (\ref{conformal1}) the perturbations $W_a$, $a_a$,$X_a/\rho$, $Y_a/p$ and $\mathcal{C}$ are also frame independent. We will come back to this problem in the next section. 

Of course we can also construct other invariant perturbations in a similar way. We will not exhaust all the possibilities in this paper, 
but we have seen that the covariant approach provides a convenient way to identify the cosmological perturbations which are both gauge and conformal invariant. 
In the next section we will see what the forms these conformal invariant perturbations have in the coordinate approach.

\section{links to the coordinate approach}

In the coordinate approach, we first make the background-perturbation splitting in a given coordinate chart. The perturbed metric has the following form
\bea
& &g_{00}=-a^2(1+2A),~~g_{0i}=a^2B_i,\nonumber\\
& &g_{ij}=a^2[(1-2\psi) \gamma_{ij}+2E_{|ij}+F^V_{i|j}+F^V_{j|i}+2h^T_{ij}]~,
\eea
where $a$ is the scale factor, $\gamma_{ij}$ is the metric of three space with constant curvature $K$,  and $A$, $B_i$, $\psi$, $E$, $F^V_{i}$ and $h^T_{ij}$ are perturbations. 
Among them $F^V_i$ is transverse, i.e., $F^{V|i}_i=0$, and $h^T_{ij}$ is transverse and traceless, i.e., $h^{T|i}_{ij}=0$ and $\gamma^{ij} h^T_{ij}=0$. 
 The notation $|i$ denotes the covariant derivative associated with the three space metric $\gamma_{ij}$. In the rest of this paper we will also use this metric to raise and lower the spatial indices. 
Furthermore, we can decompose $B_i$ into the transverse and longitudinal parts
\be
B_i=\partial_iB+B^V_i~, {\rm with }~B^{V|i}_i=0~.
\ee
According to the normalization $u_au^a=-1$, the expansion of the four-velocity to linear order is 
\be
u^0=\frac{1-A}{a},~~u^i=\frac{v^i}{a}~,
\ee
where $v^i$ is considered as a linear perturbation. With this we will soon get 
\be
u_0=-a(1+A), ~~u_i=a(B_i+v_i)~.
\ee
Similarly $v_i$ is decomposed into the transverse and longitudinal parts
\be
v_i=\partial_i v+v^V_i~,{\rm with }~v^{V|i}_i=0~.
\ee

Now we will use above equations to calculate the geometric quantities listed in the previous section. All of them are calculated up to the linear order. 
The vorticity tensor when expanding to the linear order is 
\be
\omega_{ij}=\frac{a}{2}[\partial_j(B_i+v_i)- \partial_i(B_j+v_j)]=\frac{a}{2}(\partial_j E^V_i- \partial_iE^V_j)~,
\ee
and other components vanishes, where we have defined $E^V_i\equiv B^V_i+v^V_i$. It only depends on vector perturbation. 
As mentioned before, $\omega_{ab}/S$ is both gauge and conformal invariant, so is the combination $\partial_j E^V_i- \partial_iE^V_j$. Due to the transverse condition satisfied by $E^V_i$, 
 we know that the vector perturbation $E^{V}_i$ itself is both gauge and conformal invariant.

For the shear tensor, we can show through calculations that the only non-vanishing component is $\sigma_{ij}$. 
The gauge and conformal invariant shear tensor is 
\bea\label{shear}
\frac{\sigma_{ij}}{S}
&=& (v+E')_{|ij}-\frac{1}{3}\gamma_{ij}\Delta (v+E')\nonumber\\
&+& \frac{1}{2}[(v^V_i+\partial_0 F^{V}_i)_{|j}+(v^V_j+\partial_0 F^{V}_j)_{|i}]\nonumber\\
&+& \partial_0 h^{T}_{ij}~,
\eea
where $\Delta$ is Laplacian associated with $\gamma_{ij}$, i.e., $\Delta f=\gamma^{ij} f_{|ij}$. 
The equation (\ref{shear}) means that the scalar perturbation $v+E'$, the vector perturbation $v^V_i+\partial_0 F^V_i$ and the tensor perturbation $h^T_{ij}$ are both gauge and conformal invariant. 
Since the conformal transformation will not affect the vector and tensor perturbations (because the conformal factor $\Omega$ is a scalar function), we will not consider vector and tensor perturbations in the rest of the paper.

The acceleration vector (neglecting the vector perturbation) is 
\be
a_0=0~,~~a_i=B_i'+v_i'+\mathcal{H}(B_i+v_i)+\partial_iA=\partial_i[(B+v)'+\mathcal{H}(B+v)+A]~.
\ee
so we obtain the gauge invariant scalar perturbation
\be\label{ai}
(B+v)'+\mathcal{H}(B+v)+A=(v+E')'+\mathcal{H}(v+E')+\Phi~,
\ee
where $\Phi\equiv A+(1/a)[a(B-E')]'$ is the gauge invariant metric perturbation introduced by Bardeen \cite{Bardeen:1980kt}.
We have seen from Eq. (\ref{shear}) that $v+E'$ is both gauge and conformal invariant. But we know $a_i$ is not conformal invariant in general and $\mathcal{H}$ varies from frame to frame. So based on Eq. (\ref{ai}) we cannot judge whether $\Phi$ is conformal invariant or not. When combining it with the change rule $\tilde{a}_i=a_i+D_i \ln \Omega$, we can find that $\Phi$ is not conformal invariant, it changes as
\be
\tilde{\Phi}=\Phi+\delta\ln \Omega+\frac{\Omega'}{\Omega}(B-E')~.
\ee 

The next step is to calculate $W_a$. For this purpose we first expand the variable $\alpha$ in Eq. (\ref{alpha}) to the linear order
\be
\alpha=\ln a -\psi+\frac{1}{3}\int d\eta \Delta (v+E')~,
\ee
this gives the local scale factor,
\be\label{scalefactor}
S=a [1-\psi+\frac{1}{3}\int d\eta \Delta (v+E')]~,
\ee
and the quantity $W_a$,
\be\label{wi}
W_0=0~,~~W_i=\partial_i[\mathcal{H}(B+v)-\psi+\frac{1}{3}\int d\eta \Delta (v+E')]~.
\ee
Known from the previous section, the difference $W_a-a_a$ is both gauge and conformal invariant, its spatial component is 
\be
W_i-a_i=\partial_i[-\psi-A-(B+v)'+\frac{1}{3}\int d\eta \Delta (v+E')]=-\partial_i[\Psi+\Phi+(v+E')'-\frac{1}{3}\int d\eta \Delta (v+E')]~,
\ee
where $\Psi \equiv \psi-\mathcal{H}(B-E')$ is another gauge invariant metric perturbation introduced in Ref. \cite{Bardeen:1980kt}.
Above equation means that even though neither $\Psi$ nor $\Phi$ is conformal invariant, their sum $\Psi+\Phi$ is. 
This is consistent with the computation of the Weyl tensor, for which up to linear order (only the scalar perturbations are included) the non-vanished component of the electric part is
\cite{Bruni:1992dg,Goode:1989jt}
\be
E_{ij}=\frac{1}{2}[(\Psi+\Phi)_{|ij}-\frac{1}{3}\Delta(\Psi+\Phi)\gamma_{ij}]~,
\ee
and all the components of the magnetic part vanish.
The sum $\Psi+\Phi$ corresponds to the tidal force of gravitation and has many important applications in cosmology. For example its time dependence generates the integrated Sachs-Wolfe
effect of the cosmic microwave background radiation (CMB) \cite{Hu:1994uz}. It is considered as the potential (the Weyl potential) to produce the gravitational lensing effect \cite{Lewis:2006fu}.
It is also appeared in the frame independent expression \cite{Catena:2006bd} of the ordinary Sachs-Wolfe effect of CMB \cite{Sachs:1967er}. 
The fact that $\Psi+\Phi$ is conformal invariant but $\Psi$ or $\Phi$ alone is not has important implications in cosmology. In recent years there are lots of studies (see e.g., \cite{Zhang,Zhao}) on distinguishing general relativity from modified gravity by cosmological probes. The latter is motivated by the interpretation alternative to dark energy of the current cosmic acceleration. A parameter which first comes to one's mind is the the ratio $\eta=\Psi/\Phi$. However this parameter is frame dependent, and as noticed in these studies there are no direct probes to measure it. Hence it should be replaced by other parameters which are proportional to the Weyl potential $\Psi+\Phi$. 

Similarly we can expand the gauge and conformal invariant quantity $X_a/\rho+4W_a$ to linear order, its non-vanishing component is 
\be
\frac{X_i}{\rho}+4W_i=\partial_i[\frac{\delta\rho}{\rho}-4\psi+(\frac{\rho'}{\rho}+4\mathcal{H})(B+v)]~,
\ee
this means the combination $\mathcal{R}=\frac{\delta\rho}{\rho}-4\psi+(\frac{\rho'}{\rho}+4\mathcal{H})(B+v)$ is both gauge and conformal invairant. 
This variable is more meaningful for the universe dominated by single radiation fluid. Generally the energy-momentum tensor of matter is covariantly conserved only in the frame in which the matter is minimally coupled, and in which the continuty equation follows, 
\be
\dot\rho+\Theta (\rho+p)=0~.
\ee 
In general, this equation is not valid in other frames. However, one can check easily that for single radiation fluid, $p=\rho/3$, this equation is frame independent. So in this case, in terms of 
the background continuty equation $\rho'+4\mathcal{H}\rho=0$, the combination mentioned above is 
\be
\frac{\mathcal{R}}{4}=-\psi+\frac{\delta\rho}{3(\rho+p)}~,
\ee 
this is the curvature perturbation on uniform-density hypersurfaces, used extensively in the cosmological perturbation theory. 
The analyses of the links of $Y_a/p+4W_a$ and $X_a/\rho-Y_a/p$ to the coordiante approach can be done in similar ways.

Up to now we have shown via the covariant approach that to the linear order the vector and tensor perturbations, and the sums of the scalar perturbations $v+E'$ and $\Psi+\Phi$ and so on in the coordinate approach are both gauge and conformal invariant, whatever the conformal factor is. Hereafter we will consider the case of hypersurface-orthogonal in which $u^a$ is the normal vector to the hypersurface $\Sigma={\rm constant}$. In this case, $a_a$, $W_a$, $X_a/\rho$, $Y_a/p$ and $\mathcal{C}$ are invariant under the subset of conformal transformations $\Omega=\Omega(\Sigma)$. Firstly Eqs. (\ref{ai}) and (\ref{wi}) tell us that the following expressions are conformal invariant
\be\label{zeta}
\xi\equiv \mathcal{H}(v+E')+\Phi,~\zeta\equiv -\psi+\mathcal{H}(B+v)~,
\ee
then the expansions of $X_a/\rho$ and $Y_a/p$ give the following conformal invariant quantities
\be\label{r1r2}
\mathcal{R}_1\equiv \frac{\delta\rho}{\rho}+\frac{\rho'}{\rho}(B+v)~,~\mathcal{R}_2\equiv \frac{\delta p}{p}+\frac{p'}{p}(B+v)~.
\ee
The variable $\zeta$ is the so called comoving curvature perturbation and takes an important role in cosmological perturbation theory. 
In addition, we have the three-curvature up to the linear order,
\be
\hat R=\frac{6K}{a^2}(1+2\psi)+\frac{4}{a^2}\Delta [\psi-\mathcal{H}(B+v)]~.
\ee 
In terms of  this equation one can obtain the gauge and conformal invariant perturbation $\mathcal{C}$, i.e., 
\bea\label{c1}
\mathcal{C} &=&S^2 \hat R-6K\nonumber\\
&=&a^2 [1-2\psi+\frac{2}{3}\int d\eta \Delta (v+E')]\{\frac{6K}{a^2}(1+2\psi)+\frac{4}{a^2}\Delta [\psi-\mathcal{H}(B+v)]\}-6K\nonumber\\
&=& 4 \Delta[\psi-\mathcal{H}(B+v)+K\int d\eta (v+E')]\nonumber\\
&\equiv &4 \Delta[-\zeta+K\int d\eta (v+E')]~,
\eea
again we get the result that $\zeta$ is invariant under the subset of the conformal transformations. 

The frequently quoted example is the scalar-tensor theory with single scalar field, such as the single field inflation model,  in which
\be
u_{a}=-\frac{\nabla_{a}\phi}{\sqrt{-\nabla_b\phi\nabla^b\phi}}~.
\ee
In these models we change between the Jordan and Einstein frames by the conformal transformation with $\Omega=\Omega(\phi)$, so that $u^a$ is normal to the hypersurface $\Omega={\rm constant}$. 
Above discussions tells us that the comoving curvature perturbation $\zeta$ is frame independent. 
This was known from previous studies \cite{Catena:2006bd,Chiba:2013mha,Gong,Chiba,Prokopec,Kubota:2011re} directly used the coordinate approach. Here from the viewpoint of covariant approach, it can be understood that the gauge and frame invariances of $\zeta$ are inherited from the quantities $W_a$ or $\mathcal{C}$. 
In terms of the identifications
\be
u_i=-a\partial_i(\frac{\delta\phi}{\phi'})=a\partial_i (B+v)~,~~B+v=-\frac{\delta\phi}{\phi'}~,
\ee
we can write down the detailed form of the comoving curvature perturbation 
\be
 \zeta=-\psi+\mathcal{H}(B+v)=-\psi-\mathcal{H}\frac{\delta\phi}{\phi'}~.
 \ee
The invariant 
\be
\mathcal{R}_1=\frac{\rho'}{\rho}(\frac{\delta\rho}{\rho'}-\frac{\delta\phi}{\phi'})
\ee
is proportional to the entropy perturbation between the density and the field. Certainly $\mathcal{R}_2$ is propotional to the entropy perturbation between the pressure and the field. 
Another invariant variable $v+E'$ in this case corresponds to 
\be
v+E'=-\frac{\delta\phi}{\phi'}-(B-E')=-\frac{\delta\phi^{gi}}{\phi'}~,
\ee
where $\delta\phi^{gi}\equiv \delta\phi+\phi' (B-E')$ is the gauge invariant perturbation of the scalar field \cite{Mukhanov:1990me}. This shows that $\delta\phi^{gi}/\phi'$ used in the coordinate approach is also frame independent. 

Finally it deserves pointing out that if the foliated hypersurfaces are those with $\rho={\rm constant}$, the variables mentioned above are invariant under the conformal transformations with $\Omega=\Omega(\rho)$. For this case, $B+v=-\delta\rho/\rho'$ and 
\be
\zeta=-\psi-\mathcal{H}\frac{\delta\rho}{\rho'}
\ee
is the famous curvature perturbation on uniform-density hypersurfaces. This proves that it is frame independent if the conformal factor only depends on the energy density.

\section{gauge issue of higher order perturbations}

In this paper we have found through the covariant approach some covariantly defined perturbations, such as the quantities in Eq. (\ref{conformal1}) and $W_a$, $a_a$, $X_a/\rho$, $Y_a/p$, and $\mathcal{C}$, which have both gauge and conformal invariances. These invariances are exact. 
They themselves are small perturbations and the gauge invariances are guaranteed by the Stewart-Walker Lemma. Furthermore, we did not make any approximation in proving the conformal invariances. 
The products of these quantities are thought as higher order perturbations and the gauge and conformal invariances are self-evident.  However, sometimes we need to make expansions of these quantities by the metric and matter perturbations from the coordinate approach. In the previous section we only expanded them to the linear order. The gauge issue appears when expanding to higher orders. Taking $\mathcal{C}$ for example, 
\be\label{expansion}
\mathcal{C}=\mathcal{C}^{(1)}+\mathcal{C}^{(2)}+...,
\ee
where $\mathcal{C}^{(1)}$ was obtained in Eq. (\ref{c1}) and is both gauge and conformal invariant. Nevertheless the second order perturbation $\mathcal{C}^{(2)}$ is not gauge invariant because $\mathcal{C}^{(1)}$ does not vanish. According to the Stewart-Walker Lemma the gauge invariance of $\mathcal{C}^{(2)}$ requires both $\mathcal{C}^{(0)}$ and $\mathcal{C}^{(1)}$ have zero values \cite{Bruni:1996im}, but this condition is not satisfied here. On the other hand, the conformal invariance is not affected by the perturbative expansion,  
$\mathcal{C}^{(2)}$ and its higher order quantities are still conformal invariant. So in studying higher order perturbations, we may take the point that these quantities are obtained in a fixed gauge. For example, in the expansion (\ref{expansion}), 
all the higher order perturbations $\mathcal{C}^{(n)}$ with $n>1$ are considered as perturbations in the comoving gauge. Since $\mathcal{C}$ is associated with the comoving curvature perturbation, 
we may say that the comoving curvature perturbation in the comoving gauge is conformal invariant to the full non-linear order. This is consistent with the result of \cite{Gong}.

\section{Conclusions}

The conformal transformation or Weyl rescaling is frequently used in the theories of gravity, especially the scalar-tensor theories. It transforms from one frame to another. Essentially the conformal transformation is a local field redefinitions and will not change the physics. The observables, such as the physical cosmological perturbations, should be frame independent. It is known from the studies in the coordinate approach that some key cosmological perturbations are conformal invariant. In this paper we revisited this problem via the covariant approach. We showed that in terms of this approach the cosmological perturbations which are both gauge and conformal invariant can be easily identified. We also showed that there are two kinds of invariant variables. The first kind is invariant under arbitrary conformal transformation, examples (via the coordinate approach in the linear order) include the vector and tensor perturbations, and the sums of the scalar perturbations such as $v+E'$, $\Psi+\Phi$ and so on.  The second kind is available in the case of hypersurface-orthogonal and only invariant under a subset of the conformal transformations. A famous example of this kind is the comoving curvature perturbation $\zeta$ which takes a central role in the early universe models with single scalar field. 

We should point out that the results of this paper are proved to be valid for the universe with single component. It remains to be seen whether they can be extended to the universe with multi-components.The real universe contains in general several components and the covariant and gauge invariant variables in this case have been discussed for example in Ref. \cite{Dunsby:1991xk}. It would be interesting to investigate which variables among them are conformally invariant. We leave this for the future work. 

\section{Acknowledgement}

We are grateful to the anonymous referee for useful comments and suggestions.
This work is supported in part by NSFC under Grant No. 11422543, the Fundamental Research Funds for the Central Universities, 
and the Program for New Century Excellent Talents in University.

{}


\begin{thebibliography}{}

\bibitem{Bardeen:1980kt}
  J.~M.~Bardeen,
  Phys.\ Rev.\ D {\bf 22} (1980) 1882.
  
 \bibitem{Kodama:1985bj}
  H.~Kodama and M.~Sasaki,
  Prog.\ Theor.\ Phys.\ Suppl.\  {\bf 78} (1984) 1.
  
\bibitem{Mukhanov:1990me}
  V.~F.~Mukhanov, H.~A.~Feldman and R.~H.~Brandenberger,
  Phys.\ Rept.\  {\bf 215} (1992) 203.
  
\bibitem{Ellis:1989jt}
  G.~F.~R.~Ellis and M.~Bruni,
  Phys.\ Rev.\ D {\bf 40} (1989) 1804. 
 
 \bibitem{Ellis:1989ju}
  G.~F.~R.~Ellis, J.~Hwang and M.~Bruni,
  Phys.\ Rev.\ D {\bf 40} (1989) 1819.
  
 \bibitem{Bruni:1992dg}
  M.~Bruni, P.~K.~S.~Dunsby and G.~F.~R.~Ellis,
  Astrophys.\ J.\  {\bf 395} (1992) 34. 
 
\bibitem{Dunsby:1991xk}
  P.~K.~S.~Dunsby, M.~Bruni and G.~F.~R.~Ellis,
  Astrophys.\ J.\  {\bf 395} (1992) 54. 
 
 \bibitem{Hawking:1966qi}
  S.~W.~Hawking,
  Astrophys.\ J.\  {\bf 145} (1966) 544.
  
 \bibitem{Ellis:1971pg}
  G.~F.~R.~Ellis,
  Gen.\ Rel.\ Grav.\  {\bf 41} (2009) 581
   [Proc.\ Int.\ Sch.\ Phys.\ Fermi {\bf 47} (1971) 104].
  
  \bibitem{Stewart:1974uz}
  J.~M.~Stewart and M.~Walker,
  Proc.\ Roy.\ Soc.\ Lond.\ A {\bf 341} (1974) 49.
 
\bibitem{Vitenti:2013hda}
  S.~D.~P.~Vitenti, F.~T.~Falciano and N.~Pinto-Neto,
  Phys.\ Rev.\ D {\bf 89} (2014) 10,  103538
  [arXiv:1311.6730 [astro-ph.CO]].

\bibitem{Osano:2006ew}
  B.~Osano, C.~Pitrou, P.~Dunsby, J.~P.~Uzan and C.~Clarkson,
  JCAP {\bf 0704} (2007) 003
  [gr-qc/0612108].

 \bibitem{Catena:2006bd}
  R.~Catena, M.~Pietroni and L.~Scarabello,
  Phys.\ Rev.\ D {\bf 76} (2007) 084039
  [astro-ph/0604492].
  
  \bibitem{Deruelle:2010ht}
  N.~Deruelle and M.~Sasaki,
  Springer Proc.\ Phys.\  {\bf 137} (2011) 247
  [arXiv:1007.3563 [gr-qc]].
  
  \bibitem{Chiba:2013mha}
  T.~Chiba and M.~Yamaguchi,
  JCAP {\bf 1310} (2013) 040
  [arXiv:1308.1142 [gr-qc]].
  
  \bibitem{Gong}
J.~-O.~Gong, J.~-c.~Hwang, W.~-I.~Park, M.~Sasaki and Y.~-S.~Song,
  JCAP {\bf 1109} (2011) 023 [arXiv:1107.1840 [gr-qc]].
  
  \bibitem{Chiba}
   T.~Chiba and M.~Yamaguchi,
  JCAP {\bf 0810} (2008) 021 [arXiv:0807.4965 [astro-ph]].
  
  \bibitem{Prokopec}
  T.~Prokopec and J.~Weenink,
  JCAP {\bf 1309} (2013) 027 [arXiv:1304.6737 [gr-qc]].

\bibitem{Kubota:2011re}
  T.~Kubota, N.~Misumi, W.~Naylor and N.~Okuda,
  JCAP {\bf 1202} (2012) 034 [arXiv:1112.5233 [gr-qc]].

  
 \bibitem{Bardeen:1983qw}
  J.~M.~Bardeen, P.~J.~Steinhardt and M.~S.~Turner,
  Phys.\ Rev.\ D {\bf 28} (1983) 679.
  
  \bibitem{Bezrukov:2007ep}
  F.~L.~Bezrukov and M.~Shaposhnikov,
  Phys.\ Lett.\ B {\bf 659} (2008) 703
  [arXiv:0710.3755 [hep-th]].
  
  \bibitem{Li}
  M.~Li,
  Phys.\ Lett.\ B {\bf 736} (2014) 488
  [arXiv:1405.0211 [hep-th]].
  
  \bibitem{Piao}
Y.~S.~Piao, arXiv:1109.4266;
arXiv:1112.3737.

\bibitem{Qiu:2012ia}
  T.~Qiu,
  JCAP {\bf 1206} (2012) 041
  [arXiv:1204.0189 [hep-ph]].


\bibitem{Domenech:2015qoa}
  G.~Domènech and M.~Sasaki,
  JCAP {\bf 1504} (2015) 04,  022
  [arXiv:1501.07699 [gr-qc]].
    
  \bibitem{other1}
  M.~Postma and M.~Volponi,
  Phys.\ Rev.\ D {\bf 90} (2014) 10,  103516
  [arXiv:1407.6874 [astro-ph.CO]].
  
  \bibitem{other2}
  L.~J\"{a}rv, P.~Kuusk, M.~Saal and O.~Vilson,
  Phys.\ Rev.\ D {\bf 91} (2015) 2,  024041
  [arXiv:1411.1947 [gr-qc]].

\bibitem{Langlois:2005ii}
  D.~Langlois and F.~Vernizzi,
  Phys.\ Rev.\ Lett.\  {\bf 95} (2005) 091303
  [astro-ph/0503416].
  
  \bibitem{Langlois:2005qp}
  D.~Langlois and F.~Vernizzi,
  Phys.\ Rev.\ D {\bf 72} (2005) 103501
  [astro-ph/0509078].
  
 \bibitem{Carroll:2004st}
  S.~M.~Carroll,
  ``Spacetime and geometry: An introduction to general relativity,''
  San Francisco, USA: Addison-Wesley (2004) 513 p

\bibitem{Goode:1989jt}
  S.~W.~Goode,
  Phys.\ Rev.\ D {\bf 39} (1989) 2882.

\bibitem{Hu:1994uz}
  W.~Hu and N.~Sugiyama,
  Astrophys.\ J.\  {\bf 444} (1995) 489
  [astro-ph/9407093].

\bibitem{Lewis:2006fu}
  A.~Lewis and A.~Challinor,
  Phys.\ Rept.\  {\bf 429} (2006) 1
  [astro-ph/0601594].

\bibitem{Sachs:1967er}
  R.~K.~Sachs and A.~M.~Wolfe,
  Astrophys.\ J.\  {\bf 147} (1967) 73
   [Gen.\ Rel.\ Grav.\  {\bf 39} (2007) 1929].
   
 \bibitem{Zhang}
  P.~Zhang, M.~Liguori, R.~Bean and S.~Dodelson,
  Phys.\ Rev.\ Lett.\  {\bf 99} (2007) 141302
  [arXiv:0704.1932 [astro-ph]].

\bibitem{Zhao}
  G.~B.~Zhao, T.~Giannantonio, L.~Pogosian, A.~Silvestri, D.~J.~Bacon, K.~Koyama, R.~C.~Nichol and Y.~S.~Song,
  Phys.\ Rev.\ D {\bf 81} (2010) 103510
  [arXiv:1003.0001 [astro-ph.CO]].
  
 \bibitem{Bruni:1996im}
  M.~Bruni, S.~Matarrese, S.~Mollerach and S.~Sonego,
  Class.\ Quant.\ Grav.\  {\bf 14} (1997) 2585
  [gr-qc/9609040]. 
  

\end{thebibliography}
\end{document}